\documentclass[useAMS,usenatbib]{mn2e}

\newcommand{\kms}{ km s$^{-1}$\xspace}

\newcommand{\nhI}{N$_{\rm H \ I}$\xspace}

\newcommand{\za}{$z_{abs}$}
\newcommand{\lala}{$\lambda\lambda$\xspace}
\newcommand\Lya{Lyman-$\alpha$\xspace}
\newcommand\five{$5\arcmin\times5\arcmin$}
\newcommand{\dv}{$\Delta v_{90}$}

\usepackage{graphics}
\usepackage{graphicx}
\usepackage{colordvi}
\usepackage{epsfig}
\usepackage{lscape}
\usepackage{rotating}
\usepackage{xspace}
\usepackage{float}
\usepackage{color}
\begin{document}

\title[Imaging sub-DLAs]{SOAR Imaging of sub-Damped Lyman-$\alpha$ systems at z$<$1.}
\author[J. D. Meiring, J.T Lauroesch et al.]{Joseph
  D. Meiring$^{1}$\thanks{Current Addres: Department of Astronomy,
    University of Massachusetts, Amherst MA, 01003} , 
James T. Lauroesch$^{1}$,
Lutz Haberzettl$^{1}$,
\newauthor  Varsha P. Kulkarni$^{2}$, C\'eline P\'eroux$^{3}$, Pushpa Khare$^{4}$, $\&$ Donald G. York$^{5,6}$ \\
$^{1}$Department of Physics and Astronomy, University of Louisville, Louisville, Ky 40292 USA\\
$^{2}$Department of Physics and Astronomy, University of South Carolina, Columbia, SC 29208, USA \\
$^{3}$Laboratoire d'Astrophysique de Marseille, OAMP, Universite Aix-Marseille $\&$ CNRS, 13388 Marseille cedex 13, France \\
$^{4}$Department of Physics, Utkal University, Bhubaneswar, 751004, India \\
$^{5}$Department of Astronomy and Astrophysics, University of Chicago, Chicago, IL 60637, USA \\ 
$^{6}$Enrico Fermi Institute, University of Chicago, Chicago, IL 60637, USA \\ }

\date{Accepted ... Received ...; in original form ...}

\pagerange{\pageref{firstpage}--\pageref{lastpage}} \pubyear{}

\maketitle

\label{firstpage}

\begin{abstract}
We present deep ground based imaging of the environments of five QSOs
that contain sub-Damped Lyman-$\alpha$ systems at z$<$1 with the SOAR
telescope and SOI camera.  We detect a clear surplus of galaxies in
these small fields, supporting the assumption that we are detecting the
galaxies responsible for the absorption systems. Assuming these galaxies are
at the redshift of the absorption line systems, we detect luminous
L$>$L$^{\star}$ galaxies for four of the five fields within 10\arcsec
of the QSO.  In contrast to previous imaging surveys
of DLA systems at these redshifts, which indicate a range of morphological
types and luminosities for the host galaxies of the systems, the galaxies
we detect in these sub-DLA fields appear to be luminous (L$\ga$L$_{\star}$). 
In the case of the absorber towards Q1009$-$0026 at \za=0.8866 we have
spectroscopic confirmation that the candidate galaxy is at the redshift of
the absorber, at an impact parameter of $\sim$35 kpc with a luminosity of
3 $\la$ L/L$^{\star} \la$ 8 depending on the magnitude of the $K$-correction.
These observations are in concordance with the view that sub-DLAs may
be more representative of massive galaxies than DLA systems. The environments of the absorbers span 
a range of types, from the inner disk of a galaxy, the periphery of a luminous galaxy, and the 
outskirts of interacting galaxies. The large impact parameters
to some of the candidate galaxies suggest that galactic outflows or tidal tails are
likely responsible for the material seen in
absorption. We find a weak correlation between \nhI
and the impact parameter at the 2$\sigma$ level, which may be
expected from the heterogeneous population of galaxies hosting the
absorption line systems and random orientation angles. In addition, we
detect a possible gravitationally lensed image of the BL-Lac object
Q0826-2230. 

\end{abstract}


\section{Introduction}

   One of the principal manners in which to study high redshift galaxies is through the use of Quasar (QSO) absorption 
line systems. These absorption line systems with the largest \nhI as measured by the \Lya $\lambda$ 1215 line are classified into
two groups, the sub-Damped \Lya systems  (sub-DLAs: 19.0 $<$ log \nhI $<$ 20.3 cm$^{-2}$) 
and Damped \Lya systems (DLAs: log \nhI $>$ 20.3 cm$^{-2}$). Although it is relatively easy 
to obtain high S/N spectra of QSOs with the large telescopes and sensitive high resolution spectrographs 
available today, much of the difficulty in interpreting the data from spectroscopic observations of QSO absorbers
is that the environment in which
the absorbing gas is residing is almost always unknown. There have been many studies aimed at imaging of
DLAs (e.g., \citealt{LeBrun97, Kul00, Kul01, War01, Rao03, Christ04, Christ05, Chen03, Chen05}). 
However, many of these studies have resulted in non-detections, or detections of dwarf or low surface brightness galaxies.
Higher redshift systems ($z > 1$) are increasingly
 difficult to detect, with only $\sim$10 percent of the host galaxies
of DLAs identified.

 The non-detection of luminous galaxies in most DLAs 
searched for, together with the low metallicities and weak metallicity evolution  determined via the absorption lines of 
undepleted elements such as S or Zn in the DLA systems (e.g. \citealt{PW02, Kul05, Mei06}) suggests that
while these objects do contain the majority of neutral gas in the Universe \citep{Wol95}, they may be tracing mainly
star formation in lower mass galaxies. Chemical evolution models of dwarf galaxies predict
low metallicity that slowly evolves until more recent times \citep{PT98}. 

    We have recently completed a survey of 31 sub-DLAs at $z<1.5$ using the Magellan Inamori Kyocera Echelle (MIKE) and 
Ultraviolet Visual Echelle Spectrograph (UVES) \citep{Mei07, Mei08, Mei09a, Mei09b, Kul07, Per06a, Per06b, Per08}. 
At odds with the typically low metallicities seen in DLA systems, several of the sub-DLAs in the sample showed near solar or 
above solar metallicity (i.e. [Zn/H]$\ga$0.0). In fact, a trend towards increasing metallicity with decreasing \nhI is seen 
over three decades in \nhI \citep{Mei09a, Kh07}. 

   Several possibilities exist to explain such a trend: 1) there is a dust obscuration bias 
that limits the number of observable metal rich DLAs to be observed as the amount of 
dust increases with metallicity \citep{Boi98, VP05}; 
2) the low H I column densities in the sub-DLA regime are a result of conversion of neutral hydrogen into molecular gas, which 
is forming stars and producing the metals; 3) the lower H I column densities are a result of increased ionisation from 
star forming regions that produce metals that enrich the surrounding areas; 
4) sub-DLAs are \emph{intrinsically} more likely to arise in massive galaxies which 
are typically more metal rich \citep{Kh07}. In all likelihood, it is possible that all of these effects are at work
together to produce the observed trend. 

 Few sub-DLAs have been directly imaged to date \citep{Chen05, Straka10, Christ05}. With this
 in mind, we have started a program to conduct follow up imaging of
sub-DLAs at $z<1.0$ to determine the nature of these systems. Here, we present deep 
SOAR imaging of the five QSO fields Q0826-2230, 
SDSS J1009-0026, SDSS J1228+1018, SDSS J1323-0021, and SDSS J1436-0051. The structure of this paper
is as follows: the observations are discussed
in $\S$ 2, notes on the individual fields are given in $\S$ 3, results of the 
observations are given in $\S$ 4, and a 
discussion is given in $\S$ 5. Spectra of the QSOs showing the absorption 
lines can be found in \citet{Mei07, Mei08, Mei09b, Per06a}.


\section{Observations and Data Reduction } \label{Sec:Obs}

\subsection{Observations}

   These data were obtained with the 4.1m Southern Observatories Astrophysics Research (SOAR) telescope located at
Cerro Pach\'on, and the Soar Optical Imager (SOI) camera during February and March 2009. Roughly 1.5 nights
were lost due to weather during the 4 night run.
The SOI camera has a pixel scale of $\sim$0.154 arcsec pix$^{-1}$ with 2$\times$2 binning, which was used 
throughout the observations. The two 2k$\times$4k detectors are arranged in a mosaic configuration with a $\sim$8 arcsec gap
between the detectors. The total field of view of this instrument is 5.25\arcmin$\times$5.25\arcmin.

Each field was observed in some combination of the Sloan Digital Sky Survey (SDSS) (\citealt{York00}) 
$g,r,i,z$ filter set. The targets in this investigation were selected mainly due to 
visibility at the SOAR telescope and the redshift of the absorber, favoring the lower redshift
systems in our prior spectroscopic sample to increase the chances of detections. 
Exposure times in each band and coordinates of the background QSO are given in 
Table \ref{Tab:Obs}. Each exposure was dithered by $\sim$10 arcsec 
to fill the gap between the detectors and minimize the effects of bad pixels. Individual 
exposure times ranged from 300 to 900 seconds, depending on the magnitude of the 
background QSO. Series of twilight sky flats were taken at the 
beginning of each night for each filter and used for the flat-field images. 

We have reduced these data with the \textsc{theli} pipeline available at 
\texttt{http://www.astro.uni-bonn.de/$\sim$mischa/theli.html} \citep{Erb05}. 
\textsc{theli} is a multipurpose pipeline configurable for many mosaic imagers. Data reduction proceeds with typical steps for
imaging data. The mosaic frames are first split, then overscan corrected and trimmed. Next, the frames are bias corrected. 
A flat field image is made from the twilight exposures taken with the same filter used for the science frames. 
The data are then processed and a superflat is created from the smoothed data and a fringe model 
is also created for fields observed in the $i$ and $z$ bands. 
The fringe model is then scaled and subtracted from the calibrated data. 

We used the \textsc{scamp} package for astrometric calibration and \textsc{swarp} for combining 
the individual frames. Both of these packages can be found at \texttt{http://terapix.iap.fr/}. 
The Sloan Digital Sky Survey (SDSS) astrometric catalog was used for all fields except for Q0826$-$2230, for
which we used the 2 Micron All Sky Survey (2MASS) for astrometric references.  
The fields were photometrically calibrated from the SDSS catalogues for the fields with multiple sources with $m<20.0$ detected in
both the SDSS images and these data.  The photometric
  zero point for the field of Q0826-2230, which is outside of the SDSS footprint was calibrated by determining the 
slope of the best fit line between the photometric zero points and airmasses of these fields as well as four others 
not published here.  \textsc{sextractor} version 2.5.0 was used for 
object identification and final catalog creation \citep{Bert96}. A 7$\times$7 convolution mask of a gaussian
PSF with FWHM of 4 pixels was used to filter the images during object detection. Only objects greater than
3$\sigma$ above the background noise were included in the final catalogs, with a minimum area of ten pixels. 
Magnitudes given in the text are isophotal, and centroid positions are windowed. The magnitudes have been corrected 
for Galactic extinction using the online calculator available at \texttt{http://www.ipac.caltech.edu/forms/calculator.html}. 
All magnitudes given here are in the AB system.

Three band colour images were also made of the frames using the prescription detailed in \citet{Lup04}. The images 
were aligned in pixel space, and a scaling was applied to each channel that attempted to match the colours of 
the objects seen in the SDSS images. 

\begin{table}
\begin{minipage}{96mm}
\caption{Table of observations. \label{Tab:Obs}}
\begin{tabular}{lllllllllllllllll}
\hline\hline
\multicolumn{5}{c}{Q0826$-$2230} \\
\multicolumn{5}{c}{$z_{em}>0.911$, $\alpha=08:26:01.5$, $\delta=-22:30:26.2$} \\
\hline
				& 	$g$	&	$r$	&	$i$	&	$z$	\\
\hline
Integration Time (s)		&	600	&	3300	&	3000	&   	5400	\\
FWHM ($\arcsec$)		&	1.0	&	1.0	&	0.7	&	1.25	\\
Limiting Mag. 			&	24.5	&	25.0	&	25.0	&	23.5	\\
\hline
\multicolumn{5}{c}{SDSS J1009$-$0026} \\
\multicolumn{5}{c}{$z_{em}=1.241$, $\alpha=10:09:30.40$, $\delta=-00:26:19.1$} \\
\hline
				& 	$g$	&	$r$	&	$i$	&	$z$	\\
\hline
Integration Time (s)		&	6900	&	3600	&	5400	&	2700	\\
FWHM \arcsec			&	0.9	&	0.8	&	1.1	&	1.1	\\
Lim. Magnitude			&	26.0	&	25.5	&	25.0	&	23.5	\\
\hline
\multicolumn{5}{c}{SDSS J1228+1018} \\
\multicolumn{5}{c}{$z_{em}=2.306$, $\alpha=12:28:36.80$, $\delta=+10:18:41.7$} \\
\hline
				& 	$g$	&	$r$	&	$i$	&	$z$	\\
\hline
Integration Time (s)		&	--	&	3600	&	3600	&	8100	\\
FWHM \arcsec			&	--	&	0.8	&	0.8	&	1.5	\\
Lim. Magnitude			&	--	&	25.0	&	24.5	&	23.0	\\
\hline
\multicolumn{5}{c}{SDSS J1323$-$0021} \\
\multicolumn{5}{c}{$z_{em}=1.388$, $\alpha=13:23:23.78$, $\delta=-00:21:55.3$} \\
\hline
				& 	$g$	&	$r$	&	$i$	&	$z$	\\
\hline
Integration Time (s)		&	2100	&	3600	&	4500	&	--	\\
FWHM \arcsec			&	0.7	&	0.8	&	0.7	&	--	\\
Lim. Magnitude			&	25.5	&	25.5	&	25.0	&	--	\\
\hline
\multicolumn{5}{c}{SDSS J1436$-$0051} \\
\multicolumn{5}{c}{$z_{em}=1.275$, $\alpha=14:36:45.03$, $\delta=-00:51:50.6$} \\
\hline
				& 	$g$	&	$r$	&	$i$	&	$z$	\\
\hline
Integration Time (s)		&	7200	&	3600	&	4800	&	3600	\\
FWHM \arcsec			&	0.8	&	0.8	&	0.6	&	1.4	\\
Lim. Magnitude			&	26.5	&	25.5	&	25.0	&	24.0	\\
\hline
\end{tabular}
\end{minipage}
\end{table}

\subsection{PSF Subtraction} 

Galaxies at small radii to the background QSO may become visible after PSF subtraction. 
With the \five field of view of the instrument, several suitable point sources are available in each image for reconstructing the
point spread function (PSF) of the instrument. The PSF did not vary significantly over the 
field of view of the instrument. Factors such as slighly different astrometric solutions and distortions 
determined for the individual frames as well as tracking jitters due to wind can lead to 
non-ideal PSFs in the final combined frames. 
We have used the \textsc{daophot} package in \textsc{IRAF} to create model PSFs 
using multiple stars in the images. The Moffat profile was used with $\beta=2.5$, which provided the 
best fitting model of the PSF compared to 
other options. A 30 pixel wide aperture was used in all cases.

   This model PSF was then used to interactively subtract the QSO in the images 
using the \textsc{IDP3} package in IDL. \textsc{IDP3} is
available at  \texttt{http://mips.as.arizona.edu/MIPS/IDP3/} for download.
The PSF is interactively shifted and scaled to optimally subtract the
object of interest. 

   In the case of Q0826$-$2230, the peak brightness of the QSO itself in the $r$ and $i$ bands was well 
above the level where the CCDs in the mosaic preform linearly. As such, the PSF needed to be slightly
over-subtracted for a proper fit in these images. Accurate PSF subtraction however was not a main goal 
of these observations, rather deep imaging in multiple bands to allow for galaxy 
colours to be determined was deemed a more important intent of these observations.


\section{Comments on the Individual Fields } \label{Sec:Fields}

\begin{figure*} 
\begin{center}
\includegraphics[angle=0, width=6.5in]{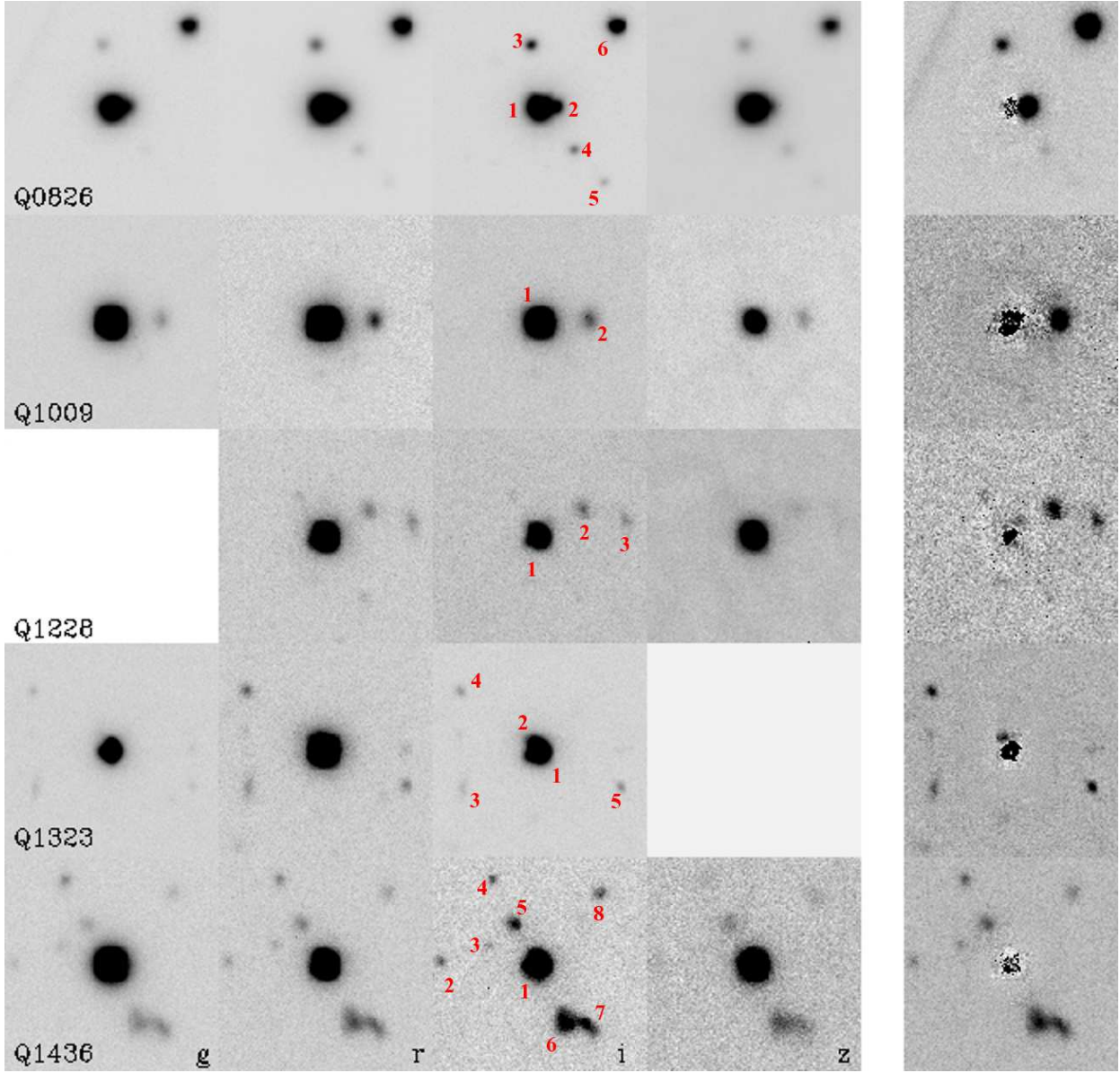}
\caption{Final coadded frames for the five QSO fields of this sample. Objects from the
 catalogues are marked in the $i$ band image. North is up and east is left in each image. 
Each thumbnail image is 20$\arcsec$ across, corresponding to $\sim$150 kpc at $z=0.8$. The 
PSF subtracted image of each field is shown in the rightmost column. \label{Fig:Fig1} }
\end{center}
\end{figure*}

\subsection{Q0826$-$2230: $z_{em}>$0.9110 \za=0.9110}

   Q0826$-$2230 is a BL-Lac object with no known redshift due to the lack of any emission lines, 
although the absorption system at \za=0.9110 does
give a lower limit to the redshift of this object \citep{Fal90, Mei07}. The sub-DLA in the spectrum of this 
object has log \nhI=19.04$\pm$0.04 (e.g. \citealt{Rao06}), 
and a high metallicity of [Zn/H]=+0.68$\pm$0.08 with a kinematical width (defined as 
the width of the profile where the inner 90 per cent of the 
absorption occurs) of \dv=278 \kms based on our earlier high resolution spectra \citep{Mei07}.

   This field was observed with the SDSS $g,r,i,z$ filters. Exposure times 
for each filter are given in Table \ref{Tab:Obs}. 
The final calibrated frames are shown in the top row of  Figure \ref{Fig:Fig1}, and 
the colour image composed of the $g,r,$ and $i$ images is shown in Figure \ref{Fig:RGB}. 
The seeing in the final combined images is $\sim$1.0, 1.0, 0.7, and 1.25 arcsec 
in the $g,r,i,$ and $z$ filters respectively. 

   Several objects are detected within 10 arcsec of the QSO, corresponding 
to $\sim$78 kpc at \za=0.9110. As this objects is near the plane of the Milky 
Way disk ($l=$243.99, $b=$+8.93), the stellar density is high in the region. 
An object 1.4 arcsec to the west of the QSO is seen most 
clearly in the $i$ band image. A Point Spread Function (PSF) model of ten 
stars near the QSO was constructed by fitting 2-d Moffat 
profiles to the individual stars and an average PSF was created that was 
used to subtract the QSO from the $i$ band image. The PSF
subtracted $i$ band image is shown in last column of Figure \ref{Fig:Fig1}.

    Limiting magnitudes of $\sim24.5,25,25$ and 23.5 were reached in the $g,r,i,z$ filters in 
the exposure times given in Table \ref{Tab:Obs}. Objects labelled four and five in Figure \ref{Fig:Fig1}
 are the only possible objects that could be galaxies at z$\sim$0.9, objects three and six 
are certainly stars. Assuming object 3  is at the redshift of the absorber, L$\ga$10L$_{\star}$ making it 
unlikely to be the host galaxy of the absorption system due to the extreme luminosity and the paucity fo such galaxies.
\textsc{hyperz} \citep{Bolz00} was used to determine photometric redshifts for objects four and five of $z_{phot}=0.66\pm0.10, 
0.62\pm0.14$ respectively.  Magnitudes for the objects are given in Table \ref{Tab:Mags}.

Object labelled number two in Figure \ref{Fig:Fig1}, located $\sim$1.5 arcsec west of the brighter object is more clearly visible in the 
PSF subtracted image shown in right column of Figure \ref{Fig:Fig1}. There are several suggestions that this QSO may be gravitationally lensed. The colours 
of this object are quite similar to the main object, the QSO has a large luminosity (M$_{i}\la-28.2$) indicative of magnification, and the side-by-side 
lensing pattern is the most commonly found lensing pattern for QSOs \citep{Koch04}. \citet{Fal90} also noted that this QSO is analogous to 
other lensed objects. Hubble Space Telescope imaging would likely reveal if this object is indeed 
gravitationally lensed.


\subsection{Q1009$-$0026: $z_{em}$=1.241 \za=0.8426,0.8866}

  There are two sub-DLAs in the spectrum of this QSO. One system has an absorption redshift of \za=0.8426  with a low metallicity 
([Zn/H]$<-$0.98 and [Fe/H]=$-1.28$ and a kinematical width of \dv=36 \kms \citet{Mei07}) and log \nhI=20.20$\pm$0.06 \citep{Rao06}, 
and a second system at \za=0.8866 has a high metallicity 
([Zn/H]=+0.25$\pm$0.06 and [Fe/H]=$-0.58$ with \dv=94 \kms \citet{Mei07}) and log \nhI=19.48 \citep{Rao06}. 

  This field was observed in the $griz$ filters. Exposure times are given in Table \ref{Tab:Obs}. The final combined frames are shown
in the second row of Figure \ref{Fig:Fig1}, and the colour combined image composed of the $gri$ frames is shown in Figure \ref{Fig:RGB}. The
seeing in the final combined frames was $\sim$0.9, 0.8, 1.1, and 1.1 arcsec in the $g,r,i,$ and $z$ filters respectively. Limiting
magnitudes of $\sim26,25.5,25,$ and 23.5 were reached in the $g,r,i,z$ filters. 

  One object is detected in the images within 10 arcsec of the QSO, 4.6 arcsec to 
the west of the QSO (corresponding to a comoving impact parameter
of $\sim$35 kpc at z=0.8426 or $\sim$36 kpc at z=0.8866). This object has been 
confirmed to be at $z=0.8866$ via integral field unit (IFU) 
observations \citep{Per10}. The \textsc{hyperz} software package 
was used to determine a photometric redshift of
$z_{phot}=0.79\pm0.08$. 

We note that photometric redshifts for star forming 
galaxies are difficult due to the flatness of their spectrum, and four band 
photometry is a minimum requirement for determining photometric redshifts. Additional photometric points in the 
Infra-red (IR) and near-UV (NUV) help to tightly constrain photometric redshifts \citep{Marg05}.  
No candidate galaxies for the \za=0.8426 system are observed in the field. 
The limiting $i$ band magnitude of $\sim25$
corresponds to a limiting luminosity of this galaxy of L$\la$0.1L$^{\star}$ in the observed frame $i$ band. 
No object at $z\sim0.8426$ was observed in the IFU observations of this QSO either \citep{Per10}.
The only other possible source in this field is a faint object south of the QSO which is minimally 
visible in the $r$ and $i$ frames, as well as the colour combined frame
in Figure \ref{Fig:RGB}. The object was not detected at $>2\sigma$ with \textsc{sextractor} however, 
and was not included in the final catalogue. A second possibility is that the 
$z=0.8426$ galaxy could be at a small impact parameter to the QSO and not have significant star formation. The 
previous IFU observations would not likely detect such a galaxy, and higher quality PSF subtraction or space based imaging 
would be necessary to rule out such a scenario.


\subsection{Q1228+1018: $z_{em}$=2.306 \za=0.9376}

   A sub-DLA system is observed in the spectrum of this QSO at \za=0.9376 with log \nhI=19.41$\pm$0.04 \citep{Rao06}. Although the 
Zn II \lala 2026,2062 lines were not detected in our earlier spectra, the system does have a high metallicity based on the
depleted element Fe of [Fe/H]=$-0.31\pm0.02$ and a kinematical width \dv=116 \kms \citep{Mei08}. 

   This field was observed in the $riz$ filters with the exposure times given in Table \ref{Tab:Obs}. The final combined frames are shown
in the third row of Figure \ref{Fig:Fig1}, and the colour combined image composed of the $riz$ frames is shown in Figure \ref{Fig:RGB}. 
The seeing in the final combined frames was $\sim$0.8\arcsec, 0.8\arcsec, and 1.5\arcsec in the $r,i,$ and $z$ filters respectively. 
Limiting magnitudes of $\sim25.0,24.5,$ and 23 were reached in the $r,i,z$ filters. 

   Two galaxies are seen northwest of the QSO at projected impact parameters of 38 and 66 kpc with $i$ band magnitudes
of 22.08 and 22.70 respectively. Two other objects are also seen in the $r$ band image of this field, but they were not 
detected above a 2$\sigma$ significance with \textsc{sextractor} and are not included in the final object catalogues.


\subsection{Q1323$-$0021: $z_{em}$=1.388 \za=0.7160}
   A metal rich sub-DLA system is observed in the spectrum of this QSO at \za=0.7160 with log \nhI=20.21$\pm$0.20 \citep{Per06a}. From high 
resolution VLT spectra of this object, metallicities based on Zn of [Zn/H]=+0.61$\pm$0.20 and based on Fe of [Fe/H]=$-0.51\pm$0.20  
and a kinematical width \dv=91 \kms  have been determined \citep{Kh04, Per06a}. 

   This field was observed in the $gri$ filters with the exposure times given in Table \ref{Tab:Obs}. The final combined frames are shown
in the fourth row of Figure \ref{Fig:Fig1}, and the colour combined image composed of the $gri$ frames is shown in Figure \ref{Fig:RGB}. 
The seeing in the final combined frames was $\sim$0.7, 0.8, and 0.7 $\arcsec$ in the $g,r,$ and $i$ filters respectively. 
Limiting magnitudes of $\sim25.5,25.5,$ and 25 were reached in the $g,r,i$ filters. 

Three objects are detected in the field, 
with one  object detected in the $i$ band image to the northeast of the QSO. 
After PSF subtraction of the QSO, the nearby object is more visible (labelled 2 in Figure \ref{Fig:Fig1}, 
and is in the same position as the object that was detected in the $K$ band IR imaging of this QSO in \citet{HW07} and 
adaptive optics imaging of \citet{Chun10}. The quality of the PSF subtraction of the QSO is not adequate 
to obtain a magnitude of this object in the $i$ band image, and we note only its position in Table \ref{Tab:Mags}. 

It is interesting to note that no H-$\alpha$ emission is detected in this galaxy \citep{Per10}. The lack of H-$\alpha$ emission and 
the  non-detections in the $g$ and $r$ images of this field (even though the seeing and depths were similar)
indicates that this galaxy is of moderately early type, with few young stars producing emission in the near UV and u band at the 
redshift of the absorber. Indeed, \citet{Per10} determine a SFR of $<$0.1M$_{\sun}$ yr$^{-1}$ for this object based on 
the non-detection of H-$\alpha$.  Magnitudes for the remaining objects are also given in Table \ref{Tab:Mags}.


\subsection{Q1436$-$0051: $z_{em}$=1.275 \za=0.7377,0.9281}

    Two systems are observed in the spectrum of this QSO at \za=0.7377 and 0.9281 with log \nhI=20.08$\pm$0.11 
and log \nhI$<$18.8 respectively. 
Both systems have high metallicities, with the lower redshift system at \za=0.7377 having [Zn/H]=$-0.05\pm$0.12 and 
[Fe/H]=$-0.61\pm$0.11 with \dv=71 \kms, while the system at 
\za=0.9281 is highly enriched with [Zn/H]$>$+0.86 and [Fe/H]$>-0.07$ and \dv=62 \kms. Both systems show 
Ca II \lala 3934 absorption lines, with [Ca/Fe]=$-0.98\pm0.03$ 
and [Ca/Fe]=$-0.79\pm0.04$ for the \za=0.7377 and \za=0.9281 systems respectively \citep{Mei08}. 

\begin{figure*} 
\includegraphics[angle=0, width=7in]{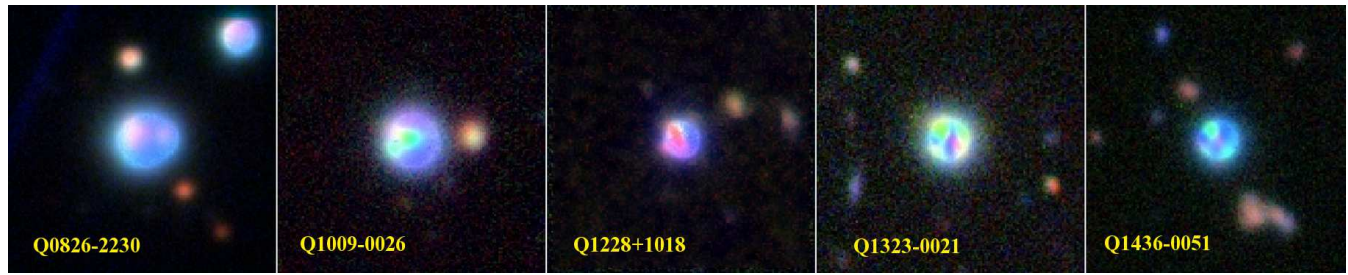}
\caption{Colour combined RGB images of the fields. In all cases, the $g,r,$ and $i$ frames where used in the RGB channels, except for Q1228+1018
for which the $r,i,$ and $z$ frames were used. \label{Fig:RGB} }
\end{figure*}

     The field of this QSO was observed in the $griz$ filters with the exposure times given in Table \ref{Tab:Obs}. The final combined frames are shown
in fifth row of Figure \ref{Fig:Fig1}, and the colour combined image composed of the $gri$ frames is shown in Figure \ref{Fig:RGB}. 
The seeing in the final combined frames was $\sim$0.8, 0.8, 0.6 and 1.4 arcsec in the $g,r,i,$ and $z$ filters respectively. 
Limiting magnitudes of $\sim26.5,25.5,25,$ and 24 were reached in the $g,r,i,z$ filters. 

     Several objects are detected within 10 arcsec of the QSO, as can be seen in Figure \ref{Fig:Fig1}. 
Objects labelled 5, 6 and 7 are the brightest galaxies in the region, with $m_i$=23.86, $m_i=20.90$ 
and $m_i=21.89$. Object 6 and 7 appear to be interacting. All of these three objects have relatively flat spectral 
energy distributions, typical of actively star forming galaxies at the redshift of the absorbers. 
Photometric redshifts for these three objects determine $z_{phot}\sim0.70$, consistent with these being members of a group. 
We note that the 1$\sigma$ error on the photometric redshift of object 5 is $\sim0.11$, i.e. the photometric redshift 
determined from \textsc{hyperz} is only  2$\sigma$ away from the redshift
 of the higher $z$ system at \za=0.9281. The extremely blue colors of objects 
labelled three and four in Figure \ref{Fig:Fig1} likely exclude these galaxies of being at the 
redshift of either absorber. \textsc{hyperz} determines photometric redshifts of 
$z_{phot}=0.25\pm0.10,0.18\pm0.10$ respectively.

\begin{table*}
\begin{tabular}{ccccccccccccccccccccccccccccccccc}
\hline\hline
QSO 		 & 	ID	 &	$\Delta\alpha$	& 	$\Delta\delta$	&	$m_g$		& 	$m_r$		&	$m_i$	 	&	$m_z$		\\
   	  	  &		&	\arcsec		&	\arcsec		&			&			&			&			\\	
\hline
Q0826$-$2230	&	1             &	--		&	--		& 16.18$\pm$0.01	&	15.74$\pm$0.01	&	15.50$\pm$0.01	&	15.36$\pm$0.01	\\
		                    &	2$^a$	  &	1.5		&	0.2		& 19.16$\pm$0.10	&	18.73$\pm$0.10	&	18.53$\pm$0.10	&	18.52$\pm$0.10	\\  
                            &	3 	           &	$-$0.8	&	6.0		& 20.90$\pm$0.05	&	29.88$\pm$0.03	&	19.51$\pm$0.03	&	19.45$\pm$0.05	\\  
                            &	4 	            &	3.2		&	$-$3.8	& 22.61$\pm$0.12	&	21.44$\pm$0.06	&	20.57$\pm$0.05	&	20.31$\pm$0.07	\\  
                            &	5 	            &	6.1		&	$-$7.1	& $>$24.50		        &	23.46$\pm$0.16	&	22.10$\pm$0.10	&	22.58$\pm$0.19	\\  
                            &	6 	            &	7.1		&	7.6		& 18.24$\pm$0.02	&	17.69$\pm$0.01	&	17.48$\pm$0.01	&	17.58$\pm$0.02	\\  

Q1009$-$0026	&	1	           &	--		&	--		& 17.97$\pm$0.02	&	17.59$\pm$0.01	&	17.63$\pm$0.02	&	17.70$\pm$0.03	\\
                           &	2	            &	4.6		&	0.1		& 22.42$\pm$0.13	&	21.86$\pm$0.10	&	21.00$\pm$0.08	&	20.83$\pm$0.12	\\

Q1228+1018	&	1  	 &	--		&	--		&	--		&	18.25$\pm$0.01	&	18.12$\pm$0.02	&	17.92$\pm$0.02	\\
                        &	2  	 &	4.1		&	2.4		&	--		&	22.93$\pm$0.12	&	22.08$\pm$0.09	&	22.00$\pm$0.15	\\  
		                &	3	     &	8.4		&	1.3		&	--		&	23.18$\pm$0.13	&	22.70$\pm$0.13	&	22.93$\pm$0.23	\\  

Q1323$-$0021	&	1	 &	--		&	--	            	&	18.24$\pm$0.01	&	17.65$\pm$0.01	&	17.41$\pm$0.01	&		--	\\
                         &	2	 &	$-$0.5$^b$	&	1.1$^b$&	--	                 	&	--	                	&	--		                &		--	\\		
		                     &3	 &	$-$7.1		&	$-$3.4		&	24.16$\pm$0.20	&	23.78$\pm$0.17	&	23.60$\pm$0.19	&		--	\\  
                             &	4	 &	$-$7.3	 	&	5.7		&	24.43$\pm$0.23	&	23.43$\pm$0.14	&	23.16$\pm$0.15	&		--	\\  
	                    	&	5	 &	7.7		    &	$-$3.4		&	25.12$\pm$0.32	&	23.87$\pm$0.17	&	23.13$\pm$0.15	&		--	\\  

Q1436$-$0051	&	1	 &	--		&	-- &	18.61$\pm$0.02	&	18.15$\pm$0.01	&	18.07$\pm$0.01	&	18.04$\pm$0.02	\\
		&	2	 &	$-$9.2		&	0.2		&	26.17$\pm$0.37	&	24.99$\pm$0.29	&	24.09$\pm$0.24	&	$>$24.00	\\  
		&	3	 &	$-$4.7		&	1.6		&	25.89$\pm$0.45	&	25.18$\pm$0.32	&	24.91$\pm$0.35	&	$>$24.00	\\  
		&	4	 &	$-$4.5		&	8.0		&	24.33$\pm$0.22	&	23.92$\pm$0.18	&	23.65$\pm$0.20	&	$>$24.00	\\  
		&	5	 &	$-$2.2		&	3.9		&	23.72$\pm$0.17	&	22.95$\pm$0.12	&	22.14$\pm$0.10	&	21.97$\pm$0.15	\\  
		&	6	 &	2.8		&	$-$5.5		&	22.49$\pm$0.10	&	21.62$\pm$0.06	&	20.82$\pm$0.05	&	20.69$\pm$0.08	\\  
		&	7	 &	4.7		&	$-$5.7		&	23.24$\pm$0.14	&	22.44$\pm$0.09	&	21.81$\pm$0.08	&	21.82$\pm$0.14	\\  
		&	8	 &	5.9		&	6.6	    	&	25.43$\pm$0.37	&	24.55$\pm$0.24	&	23.50$\pm$0.18	&	$>$24.00	\\  

\hline
\end{tabular}
\caption{AB magnitudes for the objects labeled in Figure \ref{Fig:Fig1}. The QSO itself is listed as object 1 in each field. Magnitudes given are the isophotal 
values given from \textsc{sextractor} for objects detected above a 3$\sigma$ significance. 
Limiting magnitudes are given for objects not detected in a given band. $^a$ Magnitudes for this object were determined from fitting the profile of the object after subtraction of the 
brighter object. $^b$ This object is detected in the PSF subtracted $i$ band image, but the PSF subtraction is not
adequate to determine a magnitude. \label{Tab:Mags} }
\end{table*}

\section{Results}
\subsection{Luminosities and K-corrections}
In order to determine the absolute magnitudes of the galaxies observed in these fields, $K$ corrections must be applied for an
accurate estimate. We have determined $i$-band $K$ corrections for these galaxies
assuming both a starburst type galaxy and an Sa type galaxy, with
the templates originating from the SDSS galaxy spectra template libraries available at 
\texttt{\hyphenchar\font45\relax http://www.sdss.org/dr7/algorithms/spectemplates\\/index.html}. 
Although in principle the galaxies hosting the absorbers in question could be of early type, the fact that there is significant continuum 
emission at short wavelengths apparent from the $g$ band images and flat spectral energy distributions for these galaxies indicates that star formation is currently ongoing in these 
systems. Hence, we have chosen to include only the templates for these star forming galaxies for the $K$ corrections. Assuming an early type 
galaxy spectrum would increase the absolute magnitudes of these galaxies.  
The $K$ corrected absolute magnitudes are given in Table \ref{Tab:Results}, assuming that each galaxy is at the redshift of the absorber. 
The value of M$_{i}^{\star}=-21.59$ from \citet{Blanton03} was used for determining luminosities.

\begin{figure} 
\includegraphics[angle=90, width=\linewidth]{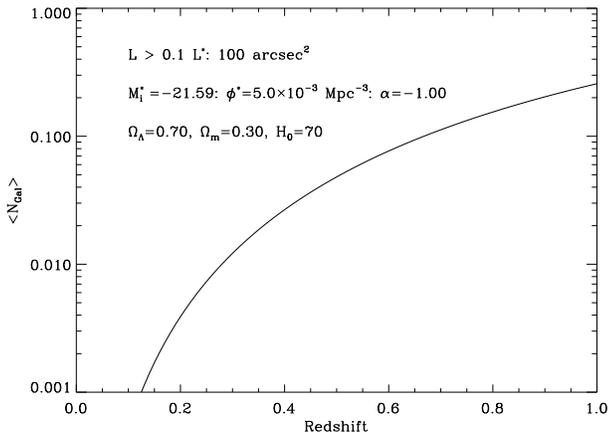}
\caption[Schecter]{The expected number of galaxies with L$>$0.1L$^{\star}$ in a pencil beam of area 100 arcsec$^2$ as a function of redshift, assuming 
the Schechter luminosity function parameters from \citet{Blanton03}.  \label{Fig:Schechter}}
\end{figure}

In total, the fields of Q1009$-$0026, Q1228+1018, Q1323$-$0021, and Q1436$-$0051 all contain $L>L^{\star}$ galaxies within 75 kpc projected impact parameters. 
Could these galaxies be unrelated field galaxies that just happen to lie near the sightline to the QSOs? This is a persistent question when imaging 
QSO absorbers, which can ultimately only be settled with followup spectroscopy to confirm the redshifts of the galaxies. However, the probability of
observing an $L>0.1L^{\star}$ galaxy in a small projected diameter pencil beam survey such as this is quite small. Assuming the Schechter function parameters 
of the $i$ band galaxy luminosity distribution from \citet{Blanton03}
of M$_{i}^{\star}=-21.59$, $\phi^{\star}=0.005$ Mpc$^{-3}$, and $\alpha=-1.00$,
we have estimated the number of galaxies with $L>0.1L^{\star}$ that will be intercepted in a random 100
arcsec$^2$ field, shown in Figure \ref{Fig:Schechter} as a function of redshift. 
As can be seen,  we expect to intercept less than one $L>0.1L^{\star}$ galaxy in a random 100 arcsec$^{2}$ field for $z<1.5$. Hence, the likelihood of each of these fields 
containing bright galaxies not associated with the hosts of the sub-DLAs are very small. 
Using the luminosity function parameters at $z\sim1$ from \citet{Ilb05} results in the same conclusions. 

\setlength{\tabcolsep}{4pt}
\begin{sidewaystable*}
\footnotesize
\begin{tabular}{cccccccccccccccccccc}
\hline\hline
Object		& \za			&  log \nhI 			&  [Zn/H]		& [Fe/H]			&	ID 	&	M$_{i}$		&	M$_{i}$		&	L$_i$/L$_i^{\star}$ & L$_i$/L$_i^{\star}$	&	$\rho$ 		\\
                                      &			& Dex				&	Dex		&	Dex			&		&	Sa		&	Burst     		&	Sa		&	Burst &    kpc		\\	
\hline
Q0826$-$2230	&  0.9110		&19.04$\pm$0.04			& +0.68$\pm$0.08	& $-0.94\pm0.06$		&	4	&	$-24.1$ 	&	$-23.2$		&	10.1	 	&	4.4		&	40.1		\\
                            &		 	            &			                          	&		                    	&			                    	&	3	&	$-25.2$	&	$-24.3$		&	27.8		&	12.1		&	47.3		\\
	                    	&			            &				                        &		                    	&				                    &	5	&	$-22.6$	& 	$-21.7$	&	2.6		&	1.1		&	72.5		\\		

Q1009$-$0026	&  0.8426, 0.8866	&20.20$\pm$0.06, 19.48$\pm$0.05	&$<-$0.98,$+0.25\pm0.06$& $-1.28\pm0.07,-0.58\pm0.06$	&	2	&	$-23.8^a$	&	$-22.7^a$	&	7.7$^a$		&	2.8$^a$		&	35.7$^a$	\\

Q1228+1018	&  0.9376		&19.41$\pm$0.02			& $<-0.37$		& $-0.30\pm0.02$		&	2	&	$-23.0$		&	$-21.9$		&	3.7		&	1.3		&	37.8		\\	
	                   	&			            &				                       &		               	&				                    &	3	&	$-22.4$		&	$-21.3$		&	2.1		&	0.8		&	66.1		\\

Q1323$-$0021	&  0.7160 		&20.21$\pm$0.19			&$+0.61\pm0.20$		& $-0.51\pm0.20$		&	2	&	--		    &	$-25.0^b$	&	--		&	6.3$^b$		&	9.0$^b$		\\
	                    	&		             	&			                        	&			                        &				                    &	3	&	$-20.3$	&	$-19.3$		&	0.3		&	0.1		&	58.4		\\
		                    &			            &				                        &			                        &			                     	&	5	&	$-20.7$		&	$-19.9$		&	0.4		&	0.2		&	59.8		\\
		                    &			            &				                        &		                            &				                    &	4	&	$-20.7$		&	$-19.9$		&	0.4		&	0.2		&	66.3		\\

Q1436$-$0051	&  0.7377, 0.9281	& 20.08$\pm$0.11, $<$18.8	& $-0.05\pm0.12,>+0.86$	& $-0.61\pm0.11,>-0.07$		&	5	&	$-21.8,-22.9$	&	$-21.0,-21.8$	&	1.2,3.4		&	0.6,1.2		&	32.7,35.2	\\
	                    	&		                     	&			                                 	&			                                    &				                                &	3	&	$-19.0,-20.1$	&	$-18.2,-19.0$	&	0.1,0.3		&	0.1,0.1		&	36.2,39.0	\\
	                       	&			                    &			                                 	&		                                    	&			                                 	&	6	&	$-23.1,-24.2$	&	$-22.3,-23.1$	&	4.0,11.0	&	1.9,4.0		&	45.0,48.4	\\
		                     &		                  	&			                                	&			                                     &			                             	&	7	&	$-22.1,-23.2$	&	$-21.3,-22.1$	&	1.6,4.4		&	0.8,1.6		&	53.9,58.0	\\
	                     	&		                     	&			                                 	&		                                     	&				                                &	8	&	$-20.4,-21.5$	&	$-19.6,-20.4$	&	0.3,0.9		&	0.2,0.3		&	64.6,69.5	\\
                            &		                    	&				                                &			                                    &				                                &	4	&	$-20.3,-21.4$	&	$-19.5,-20.3$	&	0.3,0.8		&	0.1,0.3		&	67.0,72.1	\\
                            &  	                		&			                                  	&		                                    	&				                                &	2	&	$-19.9,-21.0$	&	$-19.0,-19.8$	&	0.2,0.6		&	0.1,0.2		&	67.2,72.2	\\

\hline 
\end{tabular}
\begin{minipage}{\linewidth}{\textbf{Table 3.} Magnitudes and luminosities of the galaxies observed in the fields. \\
$^a$ As this object has been confirmed to be at the redshift of the higher $z$ absorber \citep{Per10}, the values given are for this redshift only. We estimate the limiting 
Luminosity of the \za=0.8426 system to be L$<$0.1L$^{\star}$ or the galaxy is at a small impact parameter and blended in the PSF of the QSO. \\
$^b$ Values taken from \citet{Chun10}, based on the K-band adaptive optics imaging of this system with an Sc type K-correction.  \label{Tab:Results} } 
\end{minipage}
\end{sidewaystable*}
\setcounter{table}{3}


%

\subsection{Column Density vs. Impact Parameter}

 We have compiled the available data from the literature on impact parameters for absorption systems that have 
known \nhI at $z\la1$. These data are given in Table \ref{Tab:Lit}. We have taken a sub-sample of these data
to include only the systems that have spectroscopically confirmed galaxy redshifts, or secure photometric redshifts. 
A Kendall's $\tau$ test was used to determine the probability of a correlation between 
\nhI and impact parameter $\rho$, resulting in $\tau=-0.23$ and a probability of no correlation of $0.47$.  
If one includes the data points that do not have spectroscopic redshifts of the galaxies, Kendall's $\tau$ then becomes 
$\tau=-0.48$ with a probability of no correlation of 0.05. 

We have preformed a linear regression on these data, using survival analysis to include the mixed censored data in
the fit using Schmitt's Binning method. We find the best fit trend line  of log \nhI=$(20.67\pm0.23) - (0.028\pm0.014)\times\rho$ when
using only the spectroscopically confirmed galaxies, and log \nhI=$(20.86\pm0.14) - (0.03\pm0.008)\times\rho$ when including galaxies without
redshift information.  

\begin{figure} 
\includegraphics[angle=90, width=\linewidth]{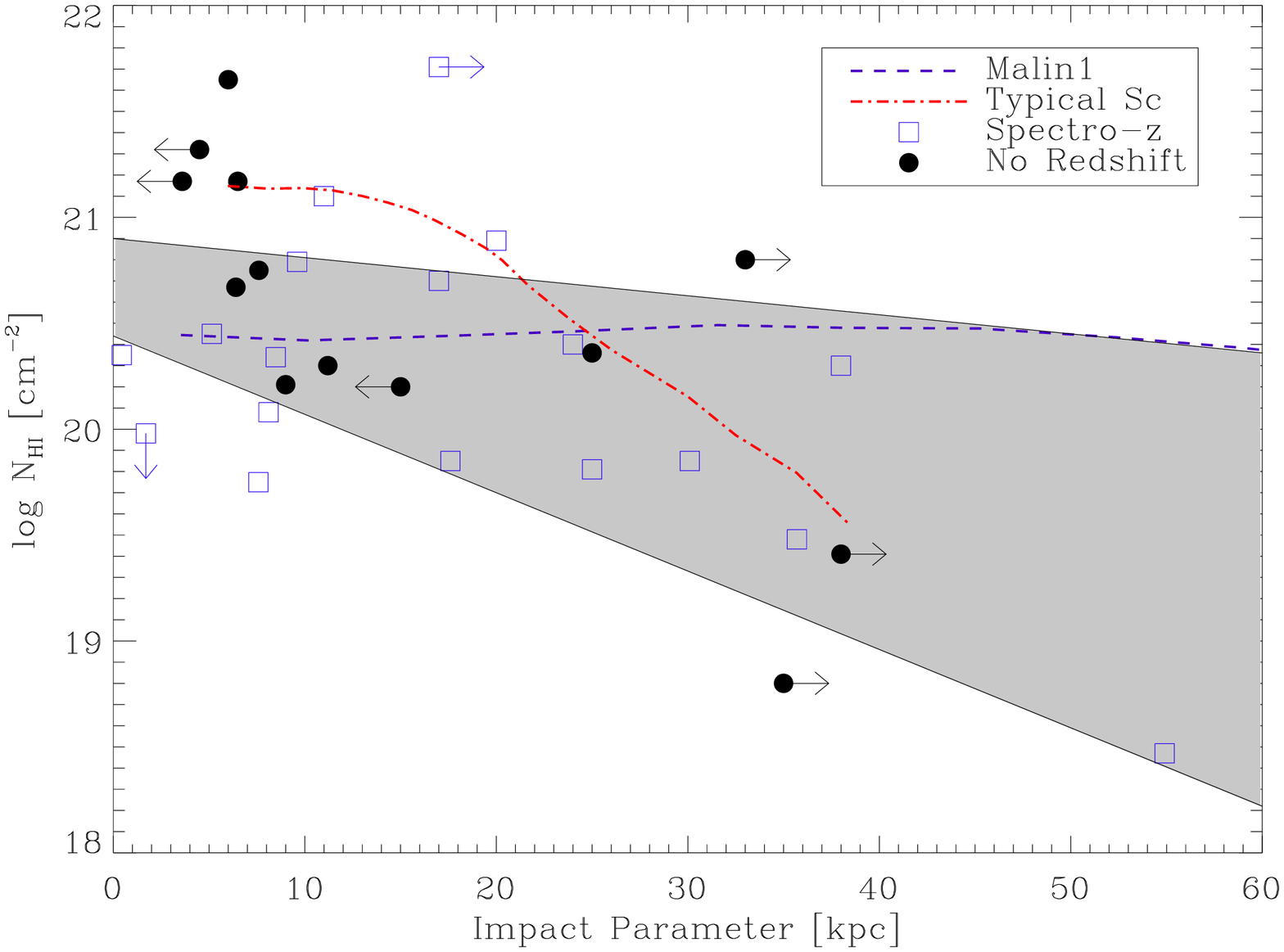}
\caption[]{Log \nhI vs impact parameter for QSO absorbers with followup imaging data. Square data points have spectroscopic or secure photometric redshift determinations of the galaxies, and
circular points have no redshift information on the galaxies. The shaded region represents the 1$\sigma$ confidence intervals on the 
best fit linear regression line using the Schmitt method to include upper and lower limits in the spectroscopically confirmed sample.  
Also overplotted is the 21 cm profile of the LSB galaxy Malin1 from \citet{Pick94}, and the average 
21 cm profile of local star forming galaxies from \citet{Swat02,Noor05}. \label{Fig:Lit} }
\end{figure}

Figure \ref{Fig:Lit} shows a plot of impact parameter vs. \nhI for the literature data, and the points from the observations 
in this investigation. For Q1228+1018, and Q1436$-$0051 we give the impact parameter as a lower limit as it is uncertain to which
galaxy in these fields the absorption is linked to without spectroscopic observations of the galaxies. 
For the absorber at \za=0.8426 in Q1009$-$0026 we give the impact parameter as an upper limit based on a 2$\arcsec$ threshold radius inside the PSF.  
Also shown in Figure \ref{Fig:Lit} are the radial 21 cm profile of Malin1, a nearby 
low surface brightness (LSB) galaxy from \citet{Pick94} and the average Sc type 21 cm radial profile from \citet{Swat02}. The data in that work are normalized relative to
the 21 cm H I radius (i.e. R/R$_H$), so we have used the median H I radius for L$\sim$L$^{\star}$ galaxies from \citet{Noor05} of 28 kpc to scale the profile. 
The minimal correlation between impact parameter and \nhI indicates that the QSO absorber population stems from a range of 
environments and morphological types, which is also supported in the heterogeneous collection of galaxies that comprise the literature sample. 

The random orientations of galaxy disk inclinations increases the dispersion in the trend
between impact parameter and \nhI. Monte Carlo simulations of $\rho$ vs \nhI reported in \citep{Chen05} have a scatter
of $\sim$0.5 dex in \nhI for a given impact parameter. These simulations were conducted using idealized smooth   
H I profiles, using more realistic simulations that include the patchy and filamentary features typically seen
 in the ISM of the Milky Way and nearby galaxies would likely increase the scatter even more. A larger sample of DLAs and sub-DLAs
with imaging data at $z<1$ would allow for a direct comparison of objects in terms of their impact parameters an 
luminosities.

\begin{table}
\begin{tabular}{lccccccccccccccccccccc} 
\hline\hline
QSO		&	\za	&	log \nhI&	$\rho$		&	$z_{gal}$ type$^a$  &	Reference	\\
		&		&	cm$^{-2}$& 	kpc		&		      &			\\
\hline
Ton1480		&	0.0036	&	20.34	&	8.48		&	s	&1				\\
Q1543+5921	&	0.0090	&	20.35	&	0.44		&	s	&1				\\
J1042+0748	&	0.0330	&	$<$19.98&	1.7		&	s	&2				\\
Q0738+313	&	0.0910	&	21.17	&	$<$3.6		&	n	&3				\\
Q0439$-$433	&	0.1010	&	19.75	&	7.57		&	s	&4				\\
Q0151+045	&	0.1600	&	19.85	&	17.6-30.1	&	s	&5, 6				\\
Q0850+440	&	0.1638	&	19.81	&	25		&	s	&7				\\
Q0738+313	&	0.2210	&	20.89	&	20		&	s	&3,4 				\\
Q0952+179	&	0.2390	&	21.32	&	$<$4.5		&	n	&8				\\
Q1127$-$145	&	0.3130	&	21.71	&	$>$17		&	s	&8,9				\\
Q0150$-$203	&	0.3830	&	18.47	&	54.9		&	s	&5				\\
Q1229$-$021	&	0.3950	&	20.75	&	7.6		&	n	&10, 11				\\
Q0809+483	&	0.4370	&	20.79	&	9.6		&	s	&10, 12				\\
Q0235+164	&	0.5240	&	21.65	&	6		&	n	&13				\\
Q0827+243	&	0.5250	&	20.3	&	38		&	s	&4, 8				\\
Q1629+120	&	0.5320	&	20.7	&	17		&	s	&8				\\
Q0058+0155	&	0.6130	&	20.08	&	8.1		&	s	&1				\\
Q1229+107	&	0.6330	&	20.3	&	11.2		&	n	&10				\\
Q1122$-$1649	&	0.6810	&	20.45	&	5.1		&	p	&1				\\
Q1328+307	&	0.6920	&	21.17	&	6.5		&	n	&10, 11				\\
Q1323$-$0031	&	0.7160	&	20.21	&	9		&	n	&14, 15				\\
Q1137+3907	&	0.7185	&	21.1	&	11		&	s	&16				\\	
Q1436$-$0051	&	0.7377	&	20.08	&	$>$33		&	n	&15				\\
Q0051+0041	&	0.7400	&	20.4	&	24		&	s	&16				\\
Q1009$-$0026	&	0.8426	&	20.2	&	$<$15		&	n	&15				\\
Q0454+039	&	0.8590	&	20.67	&	6.4		&	n	&10				\\
Q1009$-$0026	&	0.8866	&	19.48	&	35.7		&	s	&15, 17				\\
Q1436$-$0051	&	0.9281	&	$<$18.8	&	$>$35		&	n	&14				\\
Q1228+1018	&	0.9376	&	19.41	&	$>$38		&	n	&14				\\
Q1727+5302	&	0.9450	&	21.16	&	29.2		&	n	&18				\\
Q0302$-$223	&	1.0100	&	20.36	&	25.0		&	s	&10, 17				\\
\hline
\end{tabular} 
\caption{$^a$  n = no redshift information, p = secure photometric redshift, s = spectroscopic redshift \newline  
REFERENCES: (1) \citet{Chen03}, (2) \citet{Bor10}, (3) \citet{Turn01}, (4) \citet{Chen05}, (5) \citet{Ell05}, 
(6) \citet{Christ05}, (7) \citet{Lanz97}, (8) \citet{Rao03}, (9) \citet{Kac10},(10) \citet{LeBrun97}, (11) \citet{Steidel94}, 
(12) \citet{Ghar07}, (13) \citet{Burb96}, (14) \citet{Chun10} , (15) This Work, 
(16) \citet{Lacy03}, (17) \citet{Per10}, (18) \citet{HW07}  \label{Tab:Lit} }
\end{table}

\section{Discussion and Conclusions}

 In our previous spectroscopic observations of sub-DLAs at $z <1.5$, we have shown that this population is more metal rich than their DLA counterparts. With the 
observed correlation between luminosity and metallicity seen in \citet{Trem04}, one would suspect that the host galaxies of the metal rich DLAs in our sample
should be L$\ga$L$^{\star}$.  We have preformed deep imaging of the fields of five QSOs that harbour metal rich sub-DLAs 
in their spectrum with the SOAR telescope and the SOI camera. In four of the five fields, luminous galaxies are observed, 
and one QSO (Q0826-2230) is likely gravitationally lensed by the intervening absorbing galaxy. 

Is it possible that the candidate galaxies are directly atop the QSOs? The PSF subtraction of Q1323$-$0021 successfully 
reveals the galaxy detected in \citealt{Chun10}, and no other bright galaxies are detected in the PSF subtracted images so
we feel that this is unlikely. There are several reasons to suspect that the galaxies at large impact parameters
are in fact the host galaxies of these sub-DLAs. The ever possible scenario of a LSB galaxy at low impact parameter is unlikely, 
as the low metallicities of LSB galaxies \citep{McG94, Hab07} exclude them as possible candidates for these high metallicity 
sub-DLAs. The probability of observing any interloping field galaxy with L$>$L$_{\star}$ in a 
pencil beam survey such as this is small, as is shown in Figure \ref{Fig:Schechter}, and the photometric redshifts for
the galaxies are consistent with these galaxies being at substantial redshifts.
Aside from the case of Q1323$-$0021 in which the impact parameter is $\sim$9 kpc, in the fields of Q1009$-$0026, Q1228+1018, and 
Q1436$-$0051 the impact parameters of all the galaxies in the fields is $\ga$35 kpc.

These large impact parameters in several cases certainly raise the question as to the origins of the absorbing material. 
Could the material be ordinary interstellar gas residing in the extended disk of the host galaxies? 
The H I envelope of gas rich galaxies typically does extend
to much greater distances than the optical images suggest \citep{Walter08}, although not often to great enough 
radii to produce sub-DLA H I column densities at these impact parameters. 
Above a certain critical surface density, local pockets of gas are unstable to gravitational collapse \citep{Ken89, Tom64}. 
The low H I column densities of these sub-DLAs are well below the typical H I column density threshold of $\sim10^{20.5}$ cm$^{-2}$ of 
\citet{Ken89}, hence star formation is unlikely to occur in these regions. 
Observations have shown some amount of star formation at large galactic radii in nearby galaxies, however the star formation complexes are sparse
with a low filling factor, making a random sightline passing through such regions improbable \citep{Ferg98, Goddard10}. Merging 
galaxies do however often show high levels of star formation at large galactic radii \citep{Lar78, Smith10}.

Similarly, the abundances as derived from the nebular lines in the outer regions (R$>$R$_{25}$) of nearby late type galaxies 
are typically $\sim0.1Z_{\sun}$ even where the abundances are roughly solar in the inner regions of the galaxy.
These metallicities are  similar to that of many DLA systems, but well below the abundances we see in the sub-DLAs in 
this sample \citep{Ferg98, Bres09}. The impact parameters we see in these absorbers is several times greater than R$_{25}\sim10-15$ kpc typical of late type galaxies.

The relatively flat spectral energy distributions  for the galaxies in the fields of Q1009$-$0026 and Q1436$-$0051, for which we have four filter 
imaging are indicative of star forming galaxies. In the case of Q1436$-$0051, there is clear indications
of interacting galaxies which are typically actively forming stars. We feel that a more likely explanation for the origins of this gas in light of 
these imaging observations is that it is the signature of outflowing galactic winds \citep{Bou07, Kob07}. 
The high metallicities of these systems along with the large impact parameters suggest that this gas is the result of recent bursts
of star formation in the galaxies or possibly tidally stripped material in the interacting galaxies.  

The mean \dv of sub-DLAs as a population is $\sim$125 \kms, slightly higher than
the mean for DLAs of 103 \kms \citep{Kul10}. The sub-DLAs in this work have a slightly lower mean velocity width
than the total  sub-DLA population of \dv=107 \kms. 
The complex velocity profiles seen in some absorption line systems have often been attributed to 
merging or interacting galaxies \citep{Petit02, QRB08}. The profiles for both the systems in the spectrum of Q1436$-$0051
are not unusually broad, each extending $\sim$100 \kms. If the interacting galaxies seen to the south of this field are indeed 
responsible for the absorption line system, then at least in this case interacting galaxies surprisingly do not seem to produce 
a complex velocity structure. At smaller impact parameters the profiles may show  signs of broader velocity profiles.
The inclination of the disk to the QSO line of sight possibly plays a large part in the kinematic width of the absorption 
profiles.

It is interesting to note that although the recent compilation of \citet{Not08b} focused on DLA systems
(77 total systems were included in the sample, with only 7 sub-DLAs
and all had log \nhI$>$20.0), one of the systems with a high 
molecular fraction was in fact a sub-DLA. Similarly high molecule rich sub-DLAs have also been seen in other investigations
\citep{QRB08, Not08a}. It would be interesting to expand the number of H$_2$ measurements in sub-DLAs to see if they are 
perhaps also have high molecular fractions, and thus favoring the scenario where the lower H I column densities are due to the conversion of
gas from neutral to molecular. 

Although the number of DLAs or sub-DLAs that have been observed with high quality spectra has increased, there is still a 
very small number of galaxies that have been confirmed at the redshift of the absorber with followup 
spectroscopic measurements. Even fewer DLAs have spectra of the host galaxy of high enough S/N that can be used to determine the 
properties of the galaxy's stellar population such as the star formation rate (SFR), masses and star formation 
histories via spectral template fitting and emission line diagnostics. 

Followup spectroscopy is clearly necessary to
link the properties we see in absorption such as the metallicity, kinematics, and abundance patterns with the
properties of the stellar populations of the host galaxies. DLAs and sub-DLAs are likely to arise in a range of environments, a larger sample 
with imaging and spectroscopy of the host galaxies would reveal any potential dichotomy in the populations. 
The sub-DLA systems in this sample seem to arise in a range of environments themselves, from the inner region of a possible early type galaxy in the case 
of the sub-DLA in SDSS J1323$-$0021, from the periphery of a luminous star forming galaxy in SDSS J1009$-$0026 
and from the outskirts of an interacting pair of galaxies in
SDSS J1436$-$0051. However, in all cases we observe luminous galaxies that are the likely host galaxies of these metal rich absorbers.
A large sample of both spectroscopic and photometric
measurements of DLA and sub-DLA host galaxies will allow for studying trends between luminosity, impact parameter, and SFR.
Higher spatial resolution imaging with the HST would also allow for detailed information on the 
morphology of the absorbing galaxies.


\section*{Acknowledgments}
We thank the exceptionally helpful staff of SOAR for their assistance during the observing runs. 
The SOAR Telescope is a joint project of: Conselho Nacional de Pesquisas Cientificas e Tecnologicas CNPq-Brazil,
The University of North Carolina at Chapel Hill, Michigan State University, and the National Optical Astronomy Observatory.
IRAF is distributed by the National Optical Astronomy Observatory, which is operated by the Association of Universities for Research in Astronomy 
(AURA) under cooperative agreement with the National Science Foundation. We thank the anonymous referee for the helpful comments 
in the preparation of this manuscript. VPK acknowledges partial support from the National Science Foundation 
grant AST-0908890.

\bsp

\label{lastpage}

\end{document}